# Transformer-Based Financial Fraud Detection with Cloud-Optimized Real-Time Streaming

**Tingting Deng[1], Shuochen Bi [1.2], Jue Xiao [2]**

1 Independent Researcher, Simon Business School at University of Rochester, Chantilly, VA 20151, USA, dengtin1@gmail.com

2 Independent Researcher,D'Amore-McKim School of Business at Northeastern University,Boston，MA，02115，USA, bi.shu@northeastern.edu

3 Independent Researcher,The School of Business at University of Connecticut ,Jersey City，NJ，07302，USA, jue.xiao@uconn.edu

**Abstract.**

As the financial industry becomes more interconnected and reliant on digital systems, fraud detection systems must evolve to meet growing threats. Cloud-enabled Transformer models present a transformative opportunity to address these challenges. By leveraging the scalability, flexibility, and advanced AI capabilities of cloud platforms, companies can deploy fraud detection solutions that adapt to real-time data patterns and proactively respond to evolving threats. Using the Graph self-attention Transformer neural network module, we can directly excavate gang fraud features from the transaction network without constructing complicated feature engineering. Finally, the fraud prediction network is combined to optimize the topological pattern and the temporal transaction pattern to realize the high-precision detection of fraudulent transactions. The results of anti-fraud experiments on credit card transaction data show that the proposed model outperforms the 7 baseline models on all evaluation indicators: In the transaction fraud detection task, the average accuracy (AP) increased by 20% and the area under the ROC curve (AUC) increased by 2.7% on average compared with the benchmark graph attention neural network (GAT), which verified the effectiveness of the proposed model in the detection of credit card fraud transactions.

## 1. Introduction

The rapid evolution of cloud computing has redefined both technology and business landscapes. Cloud computing has become a cornerstone of enterprise innovation from its early focus on Infrastructure-as-a-Service (IaaS) to its current role as an enabler of advanced artificial intelligence (AI) applications. By integrating AI technologies such as generative models and Transformer architectures, cloud platforms now provide more than virtualized resources—they serve as the backbone for real-time, data-driven decision-making [1]. This convergence of AI and cloud computing has revolutionized industries, particularly in financial services, where the need for accurate and efficient fraud detection is paramount—transforming from tools to value drivers, cloud platforms like AWS SageMaker, Google Vertex AI, and Microsoft Azure Cognitive Services lower technical barriers, enabling companies to leverage advanced AI applications at scale while driving digital transformation.

---

[1] **Corresponding author**
[1.2] **Co-first author**

Among these advances, Transformer models have emerged as a key innovation, initially designed for natural language processing tasks but increasingly adapted for diverse data-intensive applications. As the financial industry becomes more interconnected and reliant on digital systems, fraud detection systems must evolve to meet growing threats. Cloud-enabled Transformer models present a transformative opportunity to address these challenges. By leveraging cloud platforms' scalability, flexibility, and advanced AI capabilities, companies can deploy fraud detection solutions that adapt to real-time data patterns and proactively respond to evolving threats[2]. This integration not only enhances detection accuracy and efficiency but also supports the broader goal of achieving intelligent, data-driven financial ecosystems.

Moreover, Transformers can be pre-trained on large, unlabeled datasets and then fine-tuned for specific fraud detection tasks. This transfer learning approach allows the models to learn general representations of the data and reduces the need for extensive labeled datasets. Transformers have the potential to outperform traditional methods and provide a more efficient and effective solution for fraud detection in various domains. Therefore, in this study, we aim to thoroughly explore the feasibility of applying Transformers to fraud detection.

**2. Related Work**

The detection of fraud in financial transactions has long been a pressing research challenge in the field of payment systems. As the use of mobile payments, digital wallets, and web-based payment methods has proliferated, accurately identifying the small proportion of fraudulent transactions amidst the vast volume of payment data has become an increasingly critical issue[3].

The academic literature has accumulated a rich body of research addressing this problem. Bolton and Hand [4] highlight the use of techniques like anomaly detection, classification, and clustering to identify fraudulent activities in financial systems. While effective for structured datasets, the authors note challenges in handling non-linear relationships and scalability. Their work remains foundational, with modern advancements like Transformer models and cloud-optimized systems addressing these limitations by enabling real-time, scalable, and accurate detection of evolving fraud patterns.

Christoph [5] proposed a machine-learning approach that highlights the use of techniques like anomaly detection, classification, and clustering to identify fraudulent activities in financial systems. While effective for structured datasets, the authors note challenges in handling non-linear relationships and scalability. Their work remains foundational, with modern advancements like Transformer models and cloud-optimized systems addressing these limitations by enabling real-time, scalable, and accurate detection of evolving fraud patterns, inspiring the focus of the current research.

Bolton and Hand **[6]** highlight the use of techniques like anomaly detection, classification, and clustering to identify fraudulent activities in financial systems. While effective for structured datasets, the authors note challenges in handling non-linear relationships and scalability. Their work remains foundational, with modern advancements like Transformer models and cloud-optimized systems addressing these limitations by enabling real-time, scalable, and accurate detection of evolving fraud patterns.

## 3. Transformer Model in Financial Time Series Forecasting

*3.1. The basic structure of the Transformer model*

The Transformer model is a deep learning architecture based on the self-attention mechanism proposed by Vaswani et al. in 2017, which is mainly used for processing sequence-to-sequence tasks. It consists of two parts: Encoder and Decoder[7].

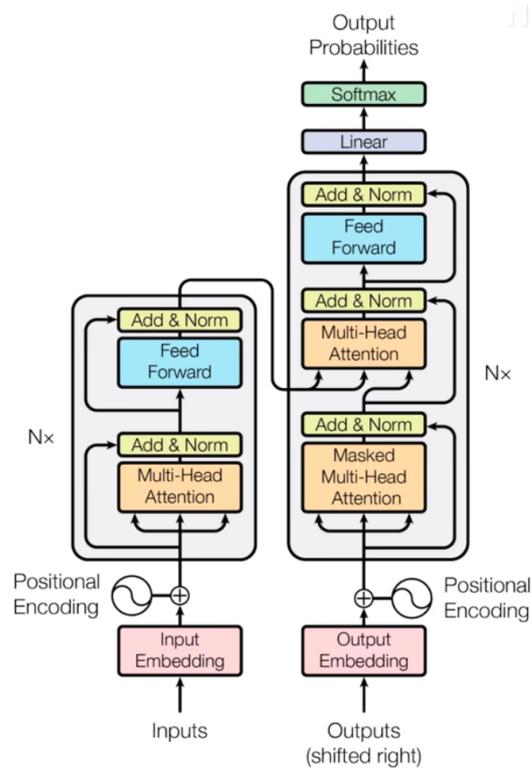

**Figure 1.** Transformer model architecture

Encoder (Encoder): The Encoder consists of multiple layers of the same structure (usually 6 layers) stacked. Each layer includes a Multi-Head Self-Attention Mechanism, and a feedforward Neural Network, Residual connections and Layer Normalization are added after each subelayer. The task of the encoder is to convert the input sequence into a potential representation.

Decoder: The structure of the decoder is similar to that of the encoder, and it is also stacked by multiple layers of the same structure (usually 6 layers)[8]. The difference is that the decoder has three sub-layers per layer: a Masked Multi-Head Self-Attention Mechanism, a multi-head self-attention mechanism, and a feedforward neural network. Through this sublayer, the decoder obtains the context information from the output of the encoder and generates the target sequence.

Self-Attention Mechanism: The self-attention mechanism is the core innovation of Transformer. It captures global dependencies by calculating how similar each element in the input sequence is to the others (i.e., the attention weight). Specifically, the input sequence is transformed linearly to generate Query, Key, and Value vectors, then computed through the dot product to get the attention weight, and then weighted to get the output.

Positional Encoding: Since the Transformer model does not contain any recursion or convolution operations, it cannot directly obtain the position information of elements in the input sequence. To complement this, Positional Encoding is added to the input embedding vector. These position codes are often generated using sine and cosine functions to ensure that the representations of different positions are unique.

*3.2. Applicability of Transformer model in financial time series forecasting*

1. Characteristics of time series data

In financial time series data, there are several important features to consider:

Non-stationary: Financial time series data is generally non-stationary, that is, its statistical properties (such as mean and variance) change over time. For example, stock prices may fluctuate dramatically because of market fluctuations or changes in economic policy.

Long-term dependencies: Financial time series data often contain long-term dependencies. For example, historical prices can have a significant impact on future prices, and changes in market trends can take a long time to manifest.

Cyclical and seasonal: Financial data may exhibit clear cyclical and seasonal patterns, such as fluctuations in stock prices when quarterly earnings are released or changes in daily trading volumes. These patterns require the model to be able to capture and model effectively.

2. How does the Transformer deal with these features

Due to its unique architecture and mechanism, the Transformer model offers several advantages when dealing with financial time series data [9]:

Self-attention mechanisms capture long-term dependencies: Self-attention mechanisms can effectively capture the interrelationships between each element in an input sequence and other elements, regardless of their distance. This means that models can better understand and capture dependencies over long time spans. For financial time series data, this can help identify long-term market trends and patterns.

Parallel processing improves computing efficiency:

Traditional RNN and LSTM networks have low computational efficiency when dealing with long sequence data because of their recursive structure. The Transformer model adopts parallel processing and can process the whole sequence of data at the same time, which greatly improves the computing efficiency. This makes it more efficient at processing large-scale financial time series data, enabling faster training and forecasting.

Location coding ensures chronological accuracy:

Because the Transformer model is disordered in nature, it cannot directly understand the order of elements in a time series. Therefore, Positional Encoding is introduced into the model to ensure that the model understands the chronological order of the input data by adding positional information to the input vector. This is important for capturing cyclical and seasonal patterns in time series data.

*3.3. Leveraging Cloud Computing for Transformer-Based Financial Fraud Detection*

1.Real-Time Data Ingestion and Processing

Cloud platforms such as AWS Kinesis and Azure Stream Analytics provide real-time data stream processing capabilities, ensuring that the Transformer model can receive data streams from financial trading systems in real time and process them quickly. This capability enables financial institutions to identify and respond to potentially fraudulent activity almost in real time as transactions occur, greatly improving the efficiency and timeliness of the model.

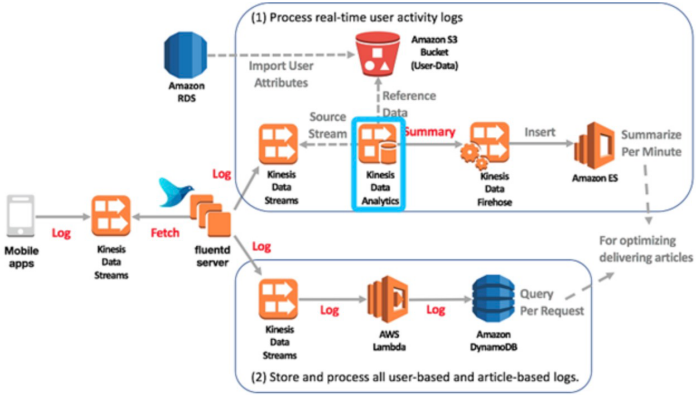

**Figure 2.** Real-time processing of user activity logs by Amazon Kinesis Data Analytics and Amazon Kinesis Data Firehose

For Example [10], AWS Kinesis can process millions of event data per second, allowing the Transformer model to process payment network and bank transaction data in real-time, and flag suspicious transactions in real-time.

2. Scalability and Elasticity

The elastic scalability of the cloud computing platform allows financial institutions to dynamically adjust computing resources to cope with fluctuations in the volume of transaction data. When needed, the cloud platform can provide large-scale computing resources for training and inference of the Transformer model to ensure that performance bottlenecks do not occur during peak periods. Example [11]: AWS SageMaker allows for distributed training with multiple GPU or TPU instances, accelerating the model training process for large-scale financial data and ensuring efficient and accurate fraud detection.

## 4. Methodology

This study introduces an innovative autoregressive model, which combines GNN and Transformer serial prediction models to detect fraudulent credit card transactions, including the overall structure of the model and the design details of each module. The model in this paper uses an innovative method to build a complex transaction graph, which can automatically learn the potential connections between transactions without complicated feature engineering, so as to detect fraud events in transaction data simply and efficiently. In addition, we have integrated hard math methods to enhance anomaly detection capabilities and improve the security and effectiveness of China's largest online payments.

*4.1. Model architecture*

This paper proposes a credit card fraud detection model (TGTN)[12] based on depth map neural network and complex transaction graph, which mainly consists of three modules (Figure 3):

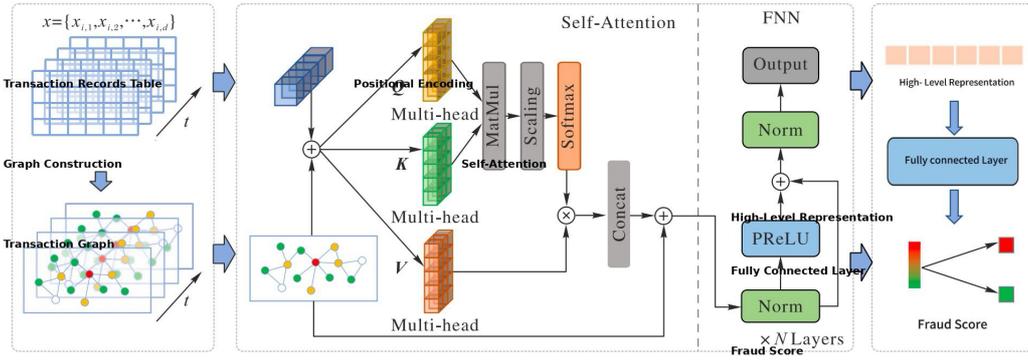

**Figure 3.** Framework of the Transaction Graph Transformer Network (TGTN) for Credit Card Fraud Detection

Data representation module: Build a complex transaction graph, and visualize the transaction data, each node represents an entity (such as users, merchants, etc.), the line represents the transaction relationship, and distinguishes different types of transactions (normal, suspicious, high-risk) by color or style, and extract the potential relationship in the transaction data.

Graph neural network module: Use deep graph neural network to model the transaction graph, capture complex transaction patterns through feature propagation aggregate high-dimensional feature representation of learning nodes, and provide accurate prediction for fraud detection.

Fraud detection output module: Analyze and visualize the detection results, highlight abnormal transactions, and present key indicators (such as accuracy, recall rate, and risk score) through the dashboard to improve the interpretability and practicality of the detection results [13].

In this paper, the model takes the original record of credit card transactions as input, without preprocessing and complex feature engineering, and outputs the fraud probability of each transaction. By automatically constructing complex transaction graphs, the model uses a graph attention Transformer to learn topological and sequential characteristics of transactions and deduces fraud probabilities through shallow prediction networks. End-to-end training is implemented as a whole to fully demonstrate the advantages of deep map learning and attention mechanisms in detecting potential fraud features.

*4.2. Build transaction graph nodes in Transformer*

The transaction graph is constructed by converting the original credit card transaction records into nodes. Each transaction node is embedded with basic transaction attributes (such as merchant number, card number, transaction amount, etc.), which are represented by the attribute matrix X. When new transaction records are generated, transaction nodes are dynamically added to the graph, and the original table structure data is transformed into a complex graph structure as input features of the Transformer network for subsequent depth map attention:

$$H(0) = \{h_1(0), h_2(0), \ldots, h_{|V|}(0)\}. \quad (1)$$
$$A \in R^{|V| \times |V|}. \quad (2)$$

The Graph Attention Transformer network enhances the node representation of the transaction graph by embedding timing information. The features of each node are updated in the graph attention layer through the following steps:

- Collection function: Collects information about neighboring nodes of a node.
- Aggregation function: Aggregate neighbor information to obtain comprehensive characteristics of nodes.
- Update function: Pass aggregate information and update node characteristics through the update function.

*4.3. Dataset description*

The experimental data set in this paper is based on partial credit card transaction records from February 1 to September 30 of a certain year. The table structure meta information of the raw transaction data is shown in Figure 3. In the experimental data, all fraud records in the above time interval were selected as negative samples. Due to the large number of credit card normal transaction records, this experiment carried out negative sampling operation for normal transactions. During the training process, data from February 1 to June 30 were selected as the training set and data from July 1 to September 30 as the test set.

During the training of the model, the parameters were selected and optimized by means of 5-fold cross-validation, and then the accuracy of the model in July, August and September was obtained by predicting the test set by month. When evaluating the model, the accuracy is calculated by comparing the prediction probability of the model with whether the actual transaction is fraudulent. The real fraudulent transaction labels in the training and testing process are derived from the user's active reporting and manual confirmation by business experts. The data set contained 141,861 transactions, of which 33,858 were fraudulent.

*4.4. Evaluation Metrics*

In this experiment, AUC (Area Under the ROC Curve) and AP (Average Precision) were used as the measurement indexes of credit card fraudulent transaction detection tasks. Through the prediction results of the model, the following evaluation indicators are obtained[14]:

**Table 1.** Model Performance Evaluation Metrics and Formulas

| Metric | Formula | Description |
| --- | --- | --- |
| AP | $AP=\frac{1}{N}\sum_{n=1}^{N}P(n)$ | Average Precision: Measures precision at each threshold |
| AUC | $AUC=\int_0^1 TPR(FPR)$ | Area Under the Curve: Measures the overall ability of the model to distinguish between classes |
| TP | True Positives | Correctly identified fraudulent transactions |
| FP | False Positives | Incorrectly identified non-fraudulent transactions as fraudulent |
| TN | True Negatives | Correctly identified non-fraudulent transactions |
| FN | False Negatives | Incorrectly identified fraudulent transactions as non-fraudulent |

*4.5. Experimental comparison model*

In order to verify the effectiveness of this model, this paper selected 7 mainstream benchmark models for comparative experiments, including SVM, RF, XGBoost, DNN, STAN and other models, and manually constructed 90 financial business features (through RFM feature engineering). For graph neural network models such as GCN, GAT and the present model, features are automatically learned directly from the transaction graph. The parameters of the benchmark model training process are set as follows:

**Table 2.** Comparison of Model Performance in Credit Card Fraud Detection Task

| Model | July AP | July AUC | August AP | August AUC | September AP | September AUC | Feature Count |
| --- | --- | --- | --- | --- | --- | --- | --- |
| SVM | 0.1203 | 0.6599 | 0.1582 | 0.6467 | 0.1172 | 0.6842 | 90 |
| RF | 0.1547 | 0.7324 | 0.1658 | 0.6841 | 0.1258 | 0.6892 | 90 |
| XGBoost | 0.2074 | 0.8894 | 0.2099 | 0.8216 | 0.257 | 0.8671 | 90 |
| DNN | 0.2512 | 0.8948 | 0.3483 | 0.9113 | 0.2509 | 0.8919 | 90 |
| STAN | 0.3021 | 0.9041 | 0.3963 | 0.9051 | 0.3315 | 0.9213 | 12 |
| GCN | 0.3473 | 0.9006 | 0.4293 | 0.8981 | 0.3275 | 0.887 | 12 |
| GAT | 0.3884 | 0.9247 | 0.4275 | 0.919 | 0.3985 | 0.9162 | 12 |
| **TGTN** | **0.4637** | **0.9471** | **0.5261** | **0.9446** | **0.4732** | **0.9429** | **12** |

Notes: **AP**: Average Precision **AUC**: Area Under the Curve

**Statistical Analysis:**

- All models were evaluated over 5 repeated experiments for July, August, and September. The table shows the average precision (AP) and AUC for each model.

- The **TGTN model** performs the best across all models, particularly in terms of AP and AUC. Compared to the other models, TGTN shows an approximately 6% improvement in AP and a 4% improvement in AUC, demonstrating its effectiveness in the credit card fraud detection task.

- This table provides a clear comparison of model performance across different months and highlights the superiority of the TGTN model.

*4.6. Experimental conclusion*

The experimental results show that the TGTN model performs well in the task of fraud detection of credit card transactions. Compared with the other 7 benchmark models, TGTN has improved in both AUC and AP indexes, among which, TGTN has improved by about 6% in AP compared with the best benchmark model and about 4% in AUC compared with GAT, verifying the effectiveness of this model. In addition, the ablation experiment shows that the location-coding module (PE) in TGTN and the self-attentional Transformer module (AT) both play an important role in the performance of the model, and TGTN-nope and TGTN-NoAT perform the same as GAT and GCN in accuracy and AUC. The effectiveness of the graph construction method and deep map learning is further proved. The model parameter sensitivity experiments show that TGTN can maintain stable performance in a large parameter space, which verifies the stability and effectiveness of the model. Finally, in the model deployment experiment, TGTN, as the core module of the credit card fraud detection system, can conduct preliminary screening through the blacklist and fraud rule engine, and conduct dynamic prediction through the transaction graph to detect and respond to new fraud patterns in real time.

**5. Conclusion**

This paper introduces a novel fraud detection model for credit card transactions based on a complex transaction graph, leveraging cloud computing and advanced deep learning techniques. The proposed model utilizes a multi-layer deep graph self-attention Transformer network to directly capture transaction graph features and sequential behavior patterns, significantly enhancing fraud detection accuracy. By constructing a complex transaction graph with transactions as nodes and their relationships as edges, our method natively incorporates temporal behavior patterns, setting it apart from traditional graph construction approaches.

Experimental results on a credit card transaction dataset demonstrate that our model outperforms seven leading benchmark models, highlighting its superior performance in fraud detection precision. Additionally, sensitivity analysis confirms the robustness and stability of the model across a wide parameter space. The deployment architecture provided for system integration into cloud-based environments ensures seamless application of the model in real-world fraud detection systems. Overall, this study provides valuable insights and practical guidelines for deploying deep graph learning methods in cloud computing environments for advanced fraud detection, offering both theoretical contributions and practical solutions.